\documentclass[twocolumn,showpacs,preprintnumbers,prl,longbibliography,fleqn,superscriptaddress]{revtex4-1}
\usepackage[english]{babel}
\usepackage[utf8]{inputenc}
\usepackage{SIunits}
\usepackage{graphicx}
\usepackage{dcolumn}
\usepackage{amsmath}
\usepackage{natbib}
\usepackage{bm}
\usepackage{textcomp}
\usepackage{color}
\usepackage{hyperref}%

\newcommand{\2}{\ensuremath{_2}}

\newcommand{\wse}{WSe\ensuremath{_2}}

\newcommand{\mose}{MoSe\ensuremath{_2}}

\newcommand{\Xzero}{X\ensuremath{_0}}
\renewcommand{\degree}{\ensuremath{^\circ\,}}
\newcommand{\Sone}{\textsf{S1}}
\newcommand{\Stwo}{\textsf{S2}}

\newcommand{\ie}{\emph{i.~e.}}

\newcommand{\dqmp}{Department of Quantum Matter Physics, University of Geneva, 24 Quai Ernest Ansermet, CH-1211 Geneva, Switzerland}
\newcommand{\gap}{Group of Applied Physics, University of Geneva, 24 Quai Ernest Ansermet, CH-1211 Geneva, Switzerland}
\newcommand{\lncmi}{Université Grenoble Alpes, INSA Toulouse, Univ. Toulouse Paul Sabatier, EMFL, CNRS, LNCMI, 38000 Grenoble, France}
\newcommand{\hbngrow}{National Institute for Materials Science, 1-1 Namiki, Tsukuba, 305-0044, Japan}

\newcommand{\lpmmc}{Université Grenoble Alpes and CNRS, LPMMC, 38042 Grenoble, France}

\definecolor{linkcol}{rgb}{0,0,0.4}
\definecolor{citecol}{rgb}{0.5,0,0}
\hypersetup{
	bookmarksopen=true,
	pdftitle="Flipping exciton angular momentum with chiral phonons in MoSe2/WSe2 heterobilayers",
	pdfauthor="Delhomme et al",
	pdfsubject="Flipping exciton angular momentum with chiral phonons in MoSe2/WSe2 heterobilayers", 
	pdfstartview={FitH},    
	pdfmenubar=true, 
	pdfhighlight=/O, 
	colorlinks=true, 
	pdfpagemode=UseNone, 
	pdfpagelayout=SinglePage, 
	pdffitwindow=true, 
	linkcolor=linkcol, 
	citecolor=linkcol, 
	urlcolor=blue
}

\begin{document}
\title{\texorpdfstring{Flipping exciton angular momentum with chiral phonons in MoSe$_2$/WSe$_2$ heterobilayers}{Flipping exciton angular momentum with chiral phonons in MoSe2/WSe2 heterobilayers}}

\author{A. Delhomme}
\affiliation{\lncmi}

\author{D. Vaclavkova}
\affiliation{\lncmi}

\author{A. Slobodeniuk}
\affiliation{\lncmi}

\author{M. Orlita}
\affiliation{\lncmi}

\author{M. Potemski}
\affiliation{\lncmi}

\author{D.M. Basko}
\affiliation{\lpmmc}

\author{K. Watanabe}
\affiliation{\hbngrow}

\author{T. Taniguchi}
\affiliation{\hbngrow}

\author{D. Mauro}   
\affiliation{\dqmp}
\affiliation{\gap}

\author{C. Barreteau}   
\affiliation{\dqmp}

\author{E. Giannini}    
\affiliation{\dqmp}

\author{A.F. Morpurgo} 
\affiliation{\dqmp}
\affiliation{\gap}

\author{N. Ubrig}   
\email{nicolas.ubrig@unige.ch}
\affiliation{\dqmp}
\affiliation{\gap}

\author{C. Faugeras}
\email{clement.faugeras@lncmi.cnrs.fr}
\affiliation{\lncmi}
\date{\today }

\begin{abstract}

Identifying quantum numbers to label elementary excitations is essential for the correct description of light-matter interaction in solids. In monolayer semiconducting transition metal dichalcogenides (TMDs) such as MoSe\2 or WSe\2, most optoelectronic phenomena are described well by labelling electron and hole states with the spin projection along the normal to the layer (S$_z$). In contrast, for WSe\2/MoSe\2 interfaces recent experiments show that taking S$_z$ as quantum number is not a good approximation, and spin mixing needs to be always considered~\cite{wang2019b,joe2019}. Here we argue that the correct quantum number for these systems is not S$_z$, but the $z$-component of the total angular momentum -- J$_z$ = L$_z$ + S$_z$ -- associated to the C$_3$ rotational lattice symmetry, which assumes half-integer values corresponding modulo 3 to distinct states. We validate this conclusion experimentally through the observation of strong intervalley scattering mediated by chiral optical phonons that -- despite carrying angular momentum 1 -- cause resonant intervalley transitions of excitons, with an angular momentum difference of 2. 

\end{abstract}


\maketitle

A useful frame to understand the optical properties of TMD monolayers is to consider spin conserving transitions, i.e. recombination of excitons with composing carriers of identical spin~\cite{wang2018}. Using spin as the quantum number to describe optical transitions in TMD monolayers, excitons have been classified as \emph{bright} for spin singlets (parallel spin configuration of electron and hole) and as \emph{dark} for spin triplets (anti-parallel spin configuration)~\cite{liu2014,Echeverry2016}. However, it has been  realized recently that the so-called \emph{dark} excitons emit light that propagates within the layer plane and can be observed using an adequate experimental configuration~\cite{molas2017,wang2017,Zhou2017}. Following this approach of spin conserving transitions, one also definitely fails to describe interlayer optical transitions in heterobilayers elaborated from distinct individual TMD monolayers~\cite{nagler2017,hsu2018}. In these systems, optical emission of interlayer exciton (IX)~\cite{gong2014,ceballos2014,rivera2015} is allowed for two particular alignment angles, i.e. $\theta = 0\degree$ or $\theta = 60\degree$~\cite{Nayak2017}. For $\theta = 0\degree$ the lowest energy exciton is a singlet state where the spin of the electron and the hole is conserved during the recombination process. The situation is different for the recombination of IX where the constituent layers have been rotated by 60$\degree$. In this case, the large extracted $g$-factor of $|g|=16$~\cite{nagler2017,seyler2019} is only compatible with a triplet state of the exciton: the involved electronic bands have anti-parallel spin configuration and one would not expect this exciton to be optically bright. This simple picture has been recently challenged by theoretical calculations~\cite{yu2018,zhang_optical_2018} that have confirmed that the excitonic triplet state in 60$\degree$ aligned heterobilayers (so called 2H-stacking~\cite{Nayak2017}) recombines radiatively.

\begin{figure*}
		\centering
		\includegraphics[width=1.0\linewidth]{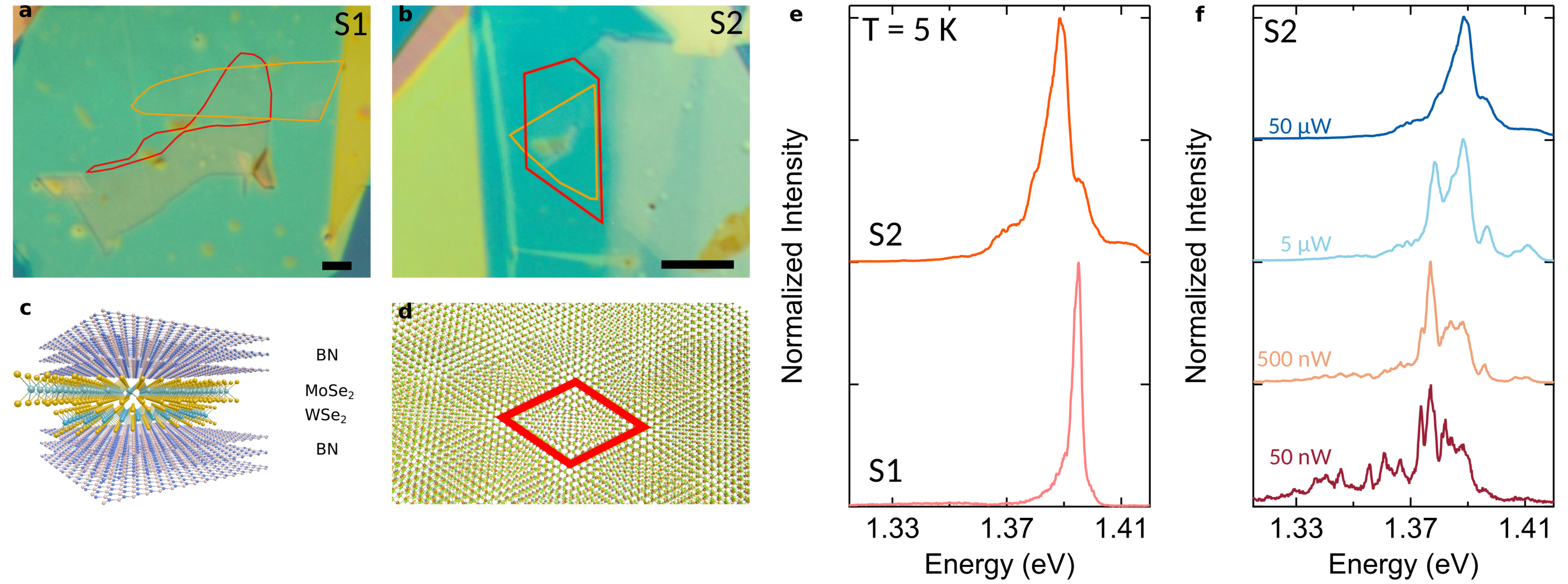}%
		\caption{\textbf{High-quality MoSe\2/WSe\2 heterobilayers with signature of the moiré potential} a,b) Optical photograph of samples \Sone\ and \Stwo\ with the TMD layers indicated by red (MoSe\2) and yellow (WSe\2) solid lines. c) Schematics of the investigated structure of an heterobilayer encapsulated in hBN. d) Moiré pattern between two TMD layers tilted by $58.5^{\circ}$ with respect to each other. e) Photoluminescence signal of sample \Sone\ and \Stwo\ in the spectral emission region of the IX. Both samples show a clear signal in this spectral range characteristic of the type-II alignment between the bands of \wse\ and \mose. FWHMs are of $4$~meV and of $15$~meV for \Sone and \Stwo, respectively  f) Photoluminescence spectra of \Stwo\ for selected laser excitation power ranging from 50~$\mu$W, to 50~nW (as indicated in the panel). The excitation wavelength of the laser is tuned to be resonant with the WSe\2\ intralayer \Xzero, \ie\ 720 nm. At the lowest exctitation power one observes the emergence of several narrow lines typical for the so-called moiré excitons. At energies slightly lower than that of the IX, a weak and broad emission can be observed (see supplementary materials for more details) which has been attributed to IX trapped in the strain field of the heterostructure~\cite{Kremser2019}.}
		\label{fig1}
	\end{figure*}

Here we show, based on magneto-PL measurements, that the quantum number in TMD monolayers is their total angular moment $J_z$ which provides a correct and extensive description of the light-matter interaction in this class of materials, including in their heterobilayers. Indeed, the dominant part of spin-orbit interaction in TMDs is proportional to $L_zS_z$ and does conserve separately $L_z$ and $S_z$, the components of the orbital and spin angular momentum, perpendicular to the layer plane. Still, a weak residual part of the spin-orbit interaction, allowed by the crystal symmetry, mixes the different spin states and only the total angular momentum $J_z=L_z+S_z$ is a good quantum number. This picture naturally explains why interlayer spin triplet excitons are optically active. These conclusions are experimentally validated by revealing a particularly efficient exciton-phonon coupling mechanism between IX in van der Waals (vdW) heterostructures made of monolayer WSe\2\ and MoSe\2, and chiral optical phonons~\cite{Zhu579} of the individual constituents, which would not be possible if spin would have been the correct quantum number of the system. This magneto-phonon resonance effect is observed when the Zeeman energy is tuned to the optical phonon energy and is responsible for an extremely efficient relaxation of excitons into the lower Zeeman component when the Zeeman energy of the IX precisely matches the energies of the non-Raman active $ E''$ optical phonon modes of WSe\2 and of MoSe\2. Finally, thanks to the finesse of the exciton-chiral phonon interaction we can further disclose that the underlying potential landscape in the heterobilayer caused by the twist angle between the monolayers affects the properties of the IX through an energy dependent $g$-factor.

\begin{figure*}
		\centering
		\includegraphics[width=1.0\linewidth]{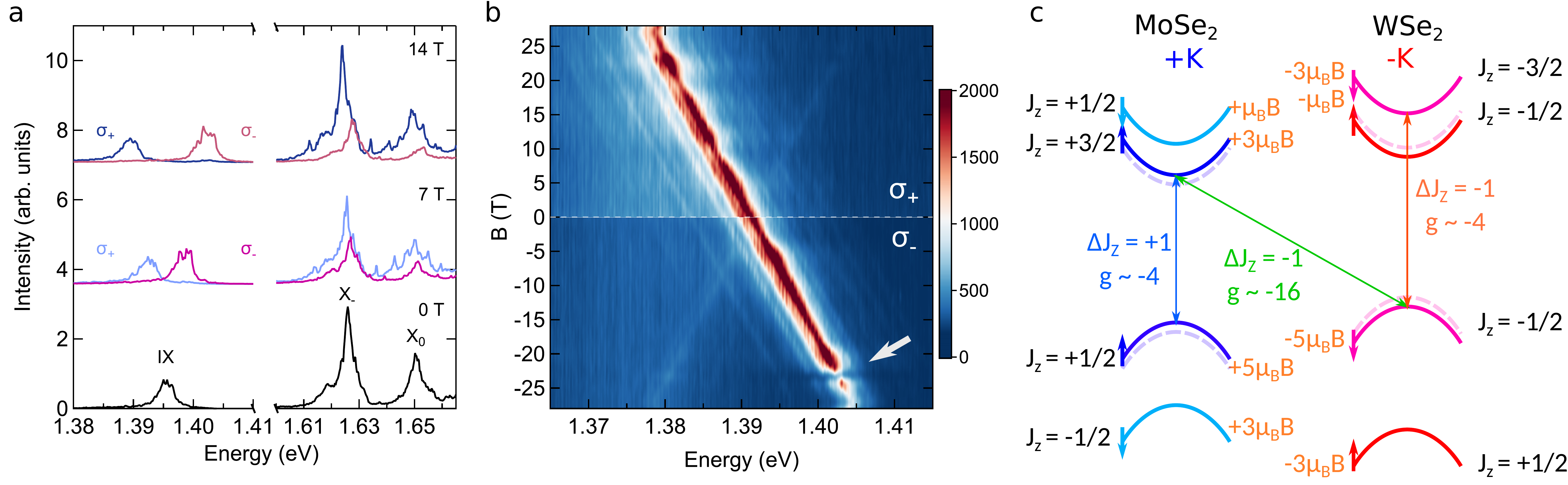}%
		\caption{\textbf{Polarization resolved Magneto-photoluminescence in high magnetic fields} a) Polarization resolved photoluminescence spectra of \Sone\ in the range of both IX and of intralayer excitons in MoSe$_2$, for selected values of the magnetic field. The laser used for the excitation is not polarized. b) False color map of polarization resolved photoluminescence intensity of the IX up to $B=\pm 29$~T. c) Schematics of the electronic bands at the $+$K point of a 60\degree aligned MoSe$_2$/WSe$_2$ heterostructure at finite magnetic field. It is formed from $\pm$K point of MoSe$_2$ and from $\mp$K points of WSe$_2$. The $-$K point of the heterostructure is obtained by reversing the spin of all the bands and changing the sign of the indicated values of $J_z$. Spin of the different bands are indicated by small colored arrows and their angular momentum J$_z$, calculated using the center of the hexagon as the center of rotation, are indicated in black. The energy shift of the bands with magnetic field is indicated in orange. The blue and red  arrows depict the intra-layer bright excitons of MoSe$_2$ and of WSe$_2$, respectively, while the green arrow is inter-layer exciton IX$_{-1}$. The change of angular momentum, $\Delta J_z$, defined modulo 3, is indicated close to the different optical excitations. This implies that $J_Z=+3/2\equiv-3/2$. The light dashed lines indicate the position of the band edges in absence of magnetic field.}
		\label{fig2}
	\end{figure*}

Two different heterobilayer samples consisting of stacked MoSe\2 and WSe\2 monolayers encapsulated in hexagonal Boron Nitride (h-BN) using a dry transfer technique~\cite{gomez2014} have been fabricated (see supplementary materials). Optical photographs are presented in Figure~\ref{fig1}a,b. The schematics of the structure and the resulting moiré pattern is illustrated by Figures~\ref{fig1}c and d, respectively. The rotation angle between the monolayers is of 60\degree, as verified below by their excitonic $g$-factor~\cite{seyler2019}. The benchmark of both samples is the presence of a peak at 1.4~eV seen in Figure~\ref{fig1}e, in addition to the emission peak of intralayer excitons of MoSe\2 and WSe\2 at higher energy~\cite{rivera2015} (for details of the experimental setup, see the Methods section at the end of the manuscript). The emission peak shown in Figure~\ref{fig1}e is the direct consequence of the type-II band alignment between MoSe\2 and WSe\2~\cite{kang2013,rivera2015,miller2017,ponomarev_semiconducting_2018}, and the experimental signature of the recombination of interlayer excitons. It is worth noting that the full width at half maximum (FWHM) of the main peak of $4$~meV (sample \Sone) is close to the intrinsic linewidth of intralayer excitons in TMD monolayers~\cite{cadiz_excitonic_2017} and the differences in the spectrum between both samples can be attributed to varying electrostatic potential landscapes. In agreement with previous reports~\cite{yu2017,seyler2019}, the IX line shape evolves when the laser power is lowered down in Figure~\ref{fig1}f and the narrow emission lines observed in the spectrum of 50~nW excitation originate from excitons trapped in the lateral moiré potential of the heterobilayer (virtually identical results are presented in the Supplementary Information for samples \Sone).

Figure~\ref{fig2}a displays the polarization resolved PL spectra of the heterostructure in sample \Sone\ for magnetic fields of $0$, $7$, and $14$~T. In agreement with previous results, the intralayer exciton in MoSe\2 are characterized by a Landé factor $g \sim -4$~\cite{Koperski2018} while a significantly larger value is found for the interlayer exciton  with $g \sim -16$. As shown in Figure~\ref{fig2}b, the Zeeman effect acting on the IX remains robust up to 30~T and all sharp features accompanying the main PL line follow the same magnetic field dependence as the bright main transition. This means that all features that we observe in this spectral region originate from the energetically most favorable region with the smallest interlayer distance~\cite{phillips2019} (AB-stacking in our case). This stacking implies that the $\pm$K points of one layer are aligned with the $\mp$K points of the other layer~\cite{Xu2014}. The large value of the $g$-factor of the interlayer exciton is fully consistent with values reported previously in the literature for 60\degree\ aligned MoSe$_2$/WSe$_2$ vdW heterostructures~\cite{nagler2017} and very recent reports have thoroughly identified that the lowest energy exciton with a $g$-factor of $g \sim -16$ belongs to an excitonic triplet state~\cite{wang2019b, joe2019}.


	\begin{figure*}
		\centering
		\includegraphics[width=0.7\linewidth]{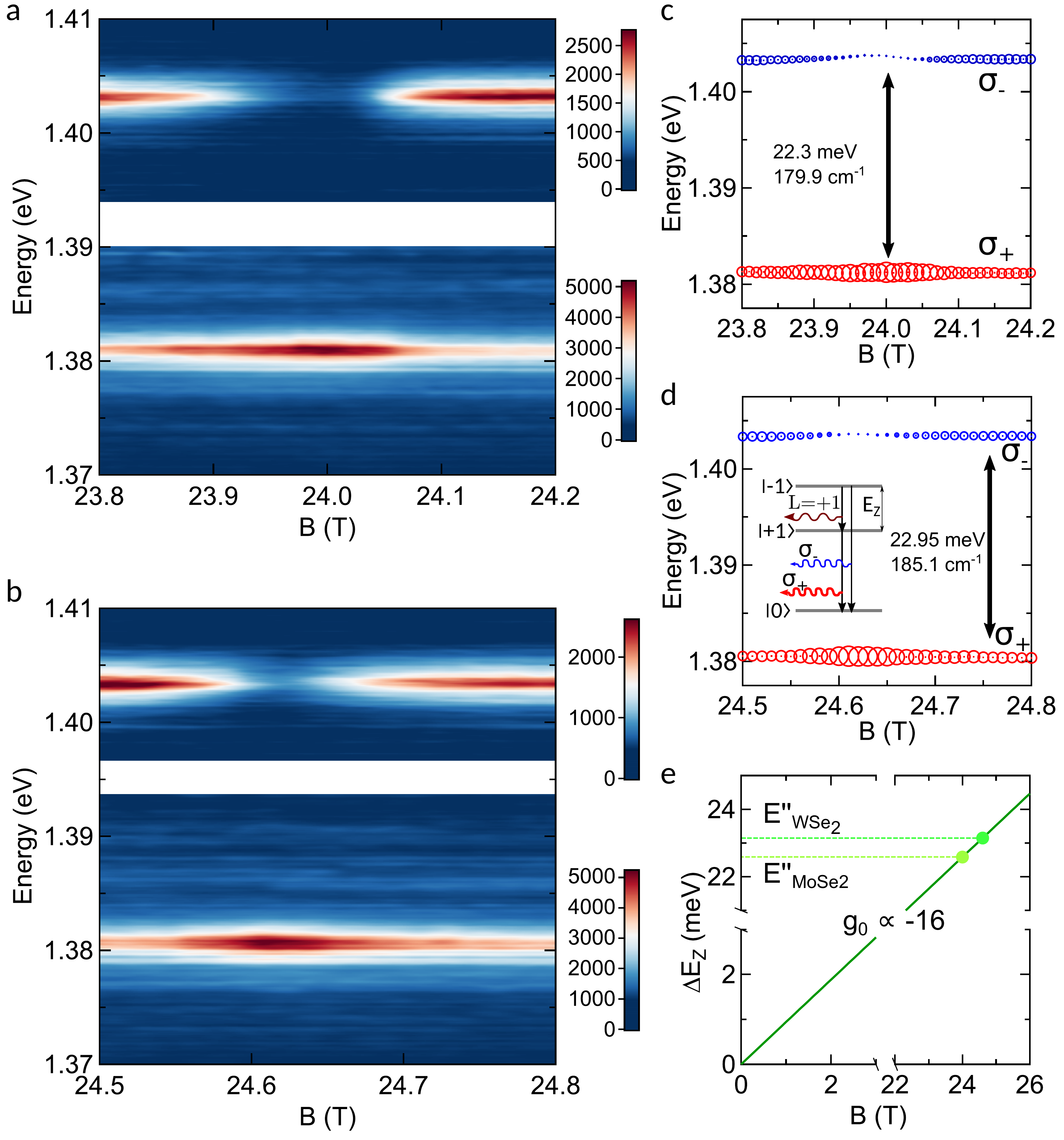}%
		\caption{\textbf{Valley relaxation due to exciton-phonon coupling} a,b) False color map of the circularly polarized PL emission from IX of \Sone\ measured with a resolution of $10$~mT in the vicinity of the two identified resonant magnetic fields. c,d) Peak position and intensity (proportional to the size of the symbol) of the polarized IX emission. The excitonic Zeeman splitting is tuned in resonance with two characteristic energies of $22.3$ and $22.95$~meV for which a very efficient thermalization of the excitons towards the lowest energy component is observed. The inset in panel d sketches the thermalization between the valleys mediated with help of a chiral phonon with L~=~+1. e) Zeeman splitting of the IX as function of the magnetic field, which shows that the energy of the chiral phonons in \wse\ and \mose\ are crossed between 24 and 25~T.}
		\label{fig3}
	\end{figure*}


In order to explain the bright recombination of the spin triplet state, we focus our attention to the schematics of the band structure presented in Figure~\ref{fig2}c. Because of the strong spin-orbit coupling in TMDs we associate to each band the total angular momentum $J_z$ following the procedure developed in References~\cite{liu2015,yu2017,yu2018}. As the AB stacked domains of the heterobilayer in which the excitonic recombination takes place are invariant under $C_3$ rotation, the angular momentum for each band is defined modulo 3. The lattice can absorb/emit units of $3\hbar$ angular momentum, which is crucial to understand the optical selection rules. We also indicate the energetic shift experienced by the different band extrema when a magnetic field is applied, by approximating the individual contributions~\cite{Aivazian2014, MacNeill2015, Srivastava2014, Stier2018, Koperski2018} to the total IX $g$-factor with $g_v \sim g_{2d} \sim 2$ and $g_{s} \sim 1$, where $g_{s}$ is the electron/hole spin, $g_v$ the valley term, and $g_{2d}$ the contribution of the atomic orbitals. Optical emission requires that the difference in angular momentum between the two bands involved in the process should be $\Delta{J_Z}\,\mathrm{modulo}\,3=\pm1$, hence the resulting photon has a helicity $\sigma\pm$. As recently noted~\cite{yu2018}, conservation of angular momentum for interlayer excitons does not require conservation of the spin during the optical transition. At the K-point of the heterostructure presented in Fig.~\ref{fig2}c, the two lowest-in-energy interlayer excitons $X_{-1}$ and $X_{+1}$, involving the $J_Z=+3/2$ and the $J_Z=+1/2$ bands in the conduction band of MoSe$_2$, respectively, with the $J_Z=-1/2$ band in the valence band of WSe$_2$, are optically active. Another example of such situation is provided by dark excitons in monolayer TMDs which carry zero total angular momentum and thus can couple to photons polarized perpendicularly to the monolayer plane~\cite{wang2017}. The experimentally observed IX in Figure~\ref{fig2}a,b appears to correspond to the IX$_{-1}$ exciton (green arrow in Figure~\ref{fig2}c), involving at the K point of the heterostructure, the lowest conduction band state of MoSe$_2$ with $J_Z=+3/2$ and the highest valence band state of WSe$_2$ with $J_Z=-1/2$. These two bands differ by $\Delta{J_z}= +2 \equiv -1$ allowing for the creation of a $\sigma-$ polarized photon, and their energy difference increases by $8\mu_B B$ when applying a magnetic field, raising a value for the factor of the IX$_{-1}$ close $g\sim -16$ (the negative sign shows that the $\sigma-$ component increases in energy under magnetic field). The IX$_{+1}$ exciton, not observed in our experiment, would be $\sigma +$ polarized with a $g$-factor close to $g\sim +12$, \ie\ showing a complete different magnetic field dependence\cite{wang2019b,joe2019}.

\begin{figure*}
		\centering
		\includegraphics[width=1.0\linewidth]{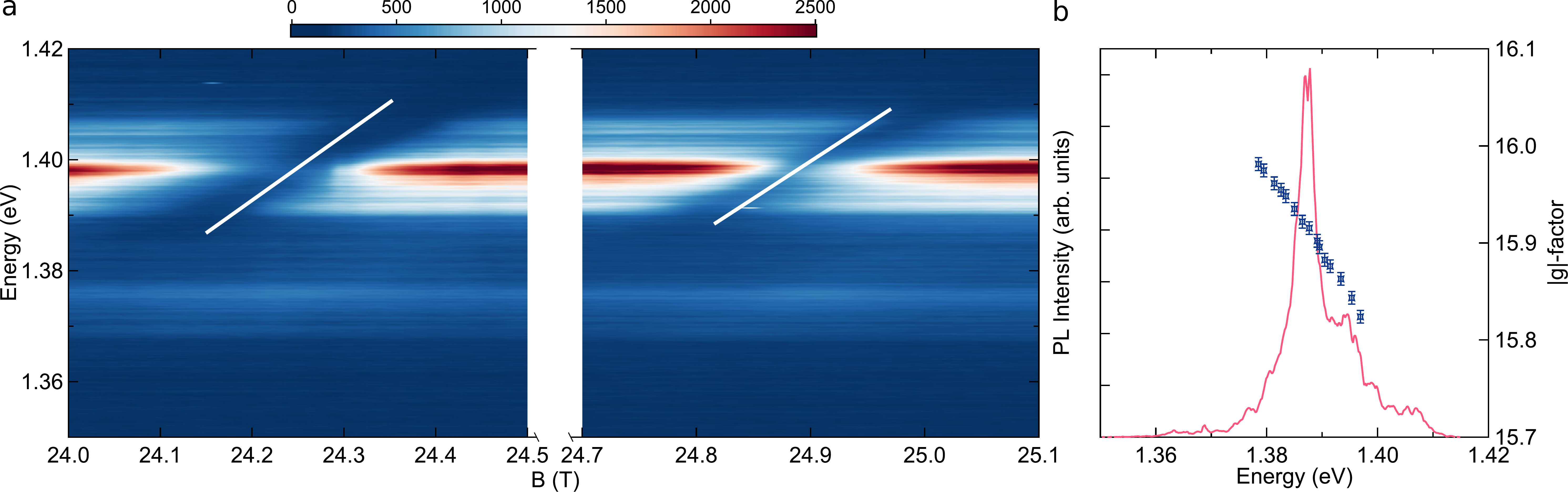}%
		\caption{\textbf{Influence of the moiré potential on the intrinsic properties of TMD heterobilayers} a) False color map of the $\sigma$- PL emission from IX of \Stwo\ measured with a resolution of 10 mT in the vicinity of the two identified resonant magnetic fields. The magnetic field for every sharp peak is different and the resonance has a \emph{diagonal} profile. The white line indicates the energy dependence of the excitonic $g$~factor, whose slope is $\partial{g}/\partial{E}\approx$ 7.5~$eV^{-1}$ (at B~$\approx$~24.2~T) and 8.5~$eV^{-1}$ (at B~$\approx$~24.9~T) for the two resonant magnetic fields. b) PL Emission spectrum (left axis) of the IX at B= 0~T of \Stwo\ with energy dependent g-factor value (right axis) extracted from the data of panel a.}
		\label{fig4}
	\end{figure*}

Optical emission from this pair of bands with opposite spin, in contrast to the case of intralayer excitons in the bare constituents, is then possible. The coupling between bands of similar angular momentum, when approaching the two layers close enough to each other, is effective both in the conduction band and in the valence band~\cite{slobodeniuk2019,gerber2019}. In a given circular polarization, the spectrum is composed of the intra-layer exciton of WSe$_2$ and of the IX at the same K point of the heterostructure, together with the intra-layer exciton of MoSe$_2$ from the -K point of the heterostructure. Magneto-optical experiments performed in Voigt geometry, up to $B=29$~T applied in the plane of the sample, did not show any evidence of magnetic brightening~\cite{molas2017,zhang2017} of any other IX transition (see Supp. Info.). These low temperature experimental data together with the recent theoretical works~\cite{yu2018} allow for the clear identification of the IX in 60\degree aligned heterobilayers of MoSe$_2$/WSe$_2$ as the IX$_{-1}$ exciton indicated in Fig.~\ref{fig2}c.

The giant $g$-factor associated with IX$_{-1}$ in $60^{\circ}$ aligned MoSe$_2$/WSe$_2$ heterobilayer allows one to tune the Zeeman energy up to few tens of meV with magnetic fields available in high magnetic fields infrastructures. In Figure~\ref{fig2}b, one can see close to $B=24$~T a nearly complete extinction of the high energy component of the IX with $\sigma -$ polarization (indicated by a white arrow). In Figure~\ref{fig3}a and \ref{fig3}b we explore this range of magnetic fields with a much higher field resolution of $10$~mT and it appears that the $\sigma -$ polarized IX emission disappears at two distinct values of the magnetic field, close to $B=24.0$~T and at $B=24.6$~T for \Sone. Simultaneously, as shown in Figure~\ref{fig3}c and \ref{fig3}d for each extinction respectively, the intensity of the lower energy $\sigma +$ component grows and decreases back to its initial value for higher magnetic fields. The excitonic emission does not show any significant energy shift nor broadening during the resonance. Its intensity vanishes in the $\sigma -$ branch at these two precise values of the magnetic field for which a very efficient relaxation of the IX is obtained. The resonant magnetic fields correspond to an excitonic Zeeman energy of $E_Z=22.30$~meV and $22.95$~meV, respectively. Striking is the narrow field interval where the energy relaxation occurs: its width of $\sim150$~mT corresponds to the change in $E_Z$ of $150~\mu$eV, as presented in Figure~\ref{fig3}a,b. This indicates a resonance with a mode with a very well defined energy, which couples IX excitons of different polarizations (i.e. from different valleys). Following our precedent discussion, such a mode reverses also the IX's angular momemtum.

A natural candidate for this mechanism is the emission of $E''$ chiral in-plane optical phonons~\cite{Zhu579}, \ie, phonons carrying an angular momentum such that $L_z=\pm1$~\cite{Zhu579,chen2018-1}. These phonons are known to efficiently couple to excitons in TMD~\cite{Song2013}, limiting their lifetime as well as the valley coherence. In particular, they produce replicas in the emission spectrum of quantum dot like structures in WSe$_2$~\cite{chen2018}. They are not Raman active in monolayers and we could not detect them in our heterobilayers (see Supplementary Information Section S6). $E''$ phonons are doubly degenerate at the $\Gamma$-point and can carry orbital angular momentum $L_z=\pm1$~\cite{Zhu579,chen2018-1} which can be transferred to the electronic system, as sketched in the inset of Figure~\ref{fig3}d. A similar situation has been also observed in the case of $E_{2g}$ phonons in graphene or in graphite~\cite{kashuba2009, kossacki2011, kossacki2012}. 
The $E''$ phonon energy at the $\Gamma$-point in TMDs, is close to $21.8$~meV in WSe$_2$~\cite{Luo2013} and to $21.1$~meV in MoSe$_2$~\cite{Soubelet2016}. These values are slightly lower than the ones observed but are known to be slightly affected by details of the experimental conditions. The high magnetic fields required to reach this resonance prevent any direct application of this effect but this very efficient scattering mechanism can play a crucial role enabling energy level lifetime engineering in optical devices.

The same phenomenon is observed in sample \Stwo\ as shown in Figure~\ref{fig4}a. Due to the slightly smaller $g$~factor for \Stwo, the resonant magnetic fields for this sample are $B=24.25$~T and $B=24.9$~T (see Figure~\ref{fig4}a). In contrast to \Sone, the resonance occurs over a broader range of  magnetic fields due to different emission components in the IX~\cite{seyler2019,tran2019}. This leads to the specific diagonal feature seen in Figure~\ref{fig4}a. The exciton-phonon resonance remains nevertheless extremely sharp for each emission peak composiing the broad IX emission. This enables to resolve the energy dependence of the excitonic $g$-factor, whose slope is $\partial{g}/\partial{E}\approx 7.5$ and $8.5$~eV$^{-1}$ for the two resonant magnetic fields, respectively. We also present in Figure~\ref{fig4}b the values of $g$-factor extracted close to $B\sim24.2$~T together with the zero field IX spectrum. The magnitude of the slope as well as the fact that the exciton $g$-factor is determined by the excitation energy are consistent with the standard band theory expression for the orbital contribution $g_i^\mathrm{orb}$ to the $g$-factor of a single-electron band~$i$ due to virtual transitions to other bands~$j$~\cite{roth59}:
\begin{equation}\label{eq:g=}
g_i^\mathrm{orb} = \sum_{j\neq i} \frac{|\langle{i}|\hat{p}_+|j\rangle|^2
-|\langle{i}|\hat{p}_-|j\rangle|^2}{m_0(\epsilon_i-\epsilon_j)},
\end{equation}
where $\hat{p}_\pm=\hat{p}_x\pm i\hat{p}_y$ are the components of the momentum
operator, $\epsilon_i,\epsilon_j$ are the band energies, and $m_0$~is the free electron mass. 
Note that the bands are predominantly localized on one of the layers, \ie\ none of the energy differences $\epsilon_i-\epsilon_j$ corresponds to the IX transition energy and therefore only an order-of-magnitude estimate can be given. Typically,  $|\langle{i}|\hat{p}_\pm|j\rangle|^2/m_0$ is several eV, a few times larger than nearest band energy differences. If different components of the luminescence spectrum are due to random band energy variations across the sample, Eq.~(\ref{eq:g=}) gives $\partial{g}/\partial{E}$ of the order of a few eV$^{-1}$, in agreement with the experimental observation in Figure~\ref{fig4}a.

To conclude, because of spin-orbit interaction, spin is not a good quantum number to describe optical transitions in monolayers of TMD and in their heterostructures. Only the total angular momentum is conserved and, because of the $C_3$ symmetry of the structure, optical transitions among conduction and valence band states with opposite spin can be allowed. Together with the ability to control the interlayer relative angle, this further provides a tool to design heterostructures with tailored electronic properties. Benefiting of the very high values for excitonic $g$-factor for 60\degree aligned heterobilayers, we demonstrate the conservation of the total angular momentum by tuning the Zeeman energy to the energy of chiral phonons and inducing a momentum transfer which flips the excitonic orbital momentum and provokes exciton intervalley scattering, polarizing entirely the excitonic system. The ability to control the lifetime of a quantum state is a key element for designing for instance emitting quantum device and using momentum transfer for this aim is a unique possibility offered by TMDs.

\section*{Acknowledgements}

\noindent We gratefully acknowledge Alexandre Ferreira for continuous and precious technical support. The work has been supported by the EC Graphene Flagship project (no. 604391), and the ANR projects ANR-17-CE24-0030 and ANR-19-CE09-0026. NU gratefully acknowledges financial support from the Swiss National Science Foundation through the Ambizione program. AFM gratefully acknowledges financial support from the Swiss National Science Foundation (Division II) and from the EU Graphene Flagship project. KW and TT acknowledge support from the Elemental Strategy Initiative conducted by the MEXT, Japan, A3 Foresight by JSPS and the CREST (JP-MJCR15F3), JST.

%

\end{document}